\documentclass[final,authoryear,3p,times,twocolumn]{elsarticle} 
\usepackage[utf8x]{inputenc} 
\usepackage{amstext,amsmath,amssymb, amsthm}
\usepackage{graphicx}
\usepackage{verbatim}
\usepackage{enumerate}

\makeatletter
\def\ps@pprintTitle{%
  \let\@oddhead\@empty
  \let\@evenhead\@empty
  \def\@oddfoot{\reset@font\hfil\thepage\hfil}
  \let\@evenfoot\@oddfoot
}
\makeatother

\begin{document}

\begin{frontmatter}

\title{Strong gender differences in reproductive success variance, \\ and the
times to the most recent common ancestors}

\author[ethz]{Maroussia~Favre\corref{cor1}}
\ead{maroussiafavre@ethz.ch, phone: +41 44 632 7166, fax: +41 44 632 1914}

\author[ethz,sfi]{Didier~Sornette}
\ead{dsornette@ethz.ch}

\address[ethz]{Department of Management, Technology and Economics, \\ Swiss Federal Institute of Technology, ETH Zurich,  Kreuzplatz 5, CH-8032 Zurich, Switzerland}
\address[sfi]{Swiss Finance Institute, c/o University of Geneva, \\ 40 blvd. Du Pont d’Arve, CH-1211 Geneva 4, Switzerland}
\cortext[cor1]{Corresponding author}

\begin{abstract}

The Time To the Most Recent Common Ancestor (TMRCA) based on human mitochondrial DNA (mtDNA) is estimated to be twice that based on the non-recombining part of the Y chromosome (NRY). These TMRCAs have special demographic implications because mtDNA is transmitted only from mother to child, while NRY is passed along from father to son. Therefore, the former locus reflects female history, and the latter, male history. To investigate what caused the two-to-one female-male TMRCA ratio $r_{F/M}=T_{F}/T_{M}$ in humans, we develop a forward-looking agent-based model (ABM) with overlapping generations. Our ABM simulates agents with individual life cycles, including life events such as reaching maturity or menopause. We implemented two main mating systems: polygynandry and polygyny with different degrees in between. In each mating system, the male population can be either homogeneous or heterogeneous. In the latter case, some males are `alphas' and others are `betas', which reflects the extent to which they are favored by female mates. A heterogeneous male population implies a competition among males with the purpose of signaling as alpha males. The introduction of a heterogeneous male population is found to reduce by a factor 2 the probability of finding equal female and male TMRCAs and shifts the distribution of $r_{F/M}$ to higher values. In order to account for the empirical observation of the factor 2, a high level of heterogeneity in the male population is needed: less than half the males can be alphas and betas can have at most half the fitness of alphas for the TMRCA ratio to depart significantly from 1. In addition, we find that, in the modes that maximize the probability of having $1.5<r_{F/M}<2.5$, the present generation has 1.4 times as many female as male ancestors. We also tested the effect of sex-biased migration and sex-specific death rates and found that these are unlikely to explain alone the sex-biased TMRCA ratio observed in humans. Our results support the view that we are descended from males who were successful in a highly competitive context, while females were facing a much smaller female-female competition.

\end{abstract}


\end{frontmatter}

\section{Introduction}

All humans alive today share common ancestors, which can be dated along specific ancestry lines using DNA samples. Human mitochondrial DNA (mtDNA) and the non-recombining part of the Y chromosome (NRY) have received particularly great attention in the context of modern human origin. While mtDNA is transmitted exclusively from mother to child (female or male), NRY is passed along only from father to son. Unexpectedly, genetic studies have shown that the Time to the Most Recent Common Ancestor (TMRCA) along all-females ancestry lines (female TMRCA or $T_{F}$) is about twice that along all-males lines (male TMRCA or $T_{M}$): 170-240 thousand years for the female TMRCA \citep{Tang, Ingman, Cann}, compared with 46-110 thousand years for the male TMRCA \citep{Tang, Pritchard, Thomson, Hammer2002}.

The finding of the ratio $r_{F/M}=T_{F}/T_{M}\simeq 2$ is important for several reasons. A first interest lies in the interpretation of the concept of effective population size. Geneticists use the Wright-Fisher model to establish a correspondence between the genetic diversity of a population, its effective population size and its TMRCA. The Wright-Fisher model assumes a haploid population (i.e. all-female or all-male), random mating, a Poisson distribution of reproductive success, non overlapping generations and a constant population size. The Wright-Fisher population size that would have yielded the same genetic diversity in the model as in the reality is defined as the effective population size. In this model, the effective population is proportional to the TMRCA. A two-to-one TMRCA ratio thus means a two-to-one effective population size. However, the concept of effective population size is merely a mathematical tool and does not correspond to a `real' population size (e.g. census or breeding size). Approximations have been developed that give a connection between effective population size and breeding size \citep{Hedrick}. A better understanding of the origin of the ratio $r_{F/M} \simeq 2$ would provide insights into the correspondence between female and male breeding populations (and underlying social structure) and TMRCA ratio. As we explain below, one contribution of the ABM that we develop below is to give a concrete realization of the correspondence between female and male breeding populations and TMRCA ratio through a representation as realistic as possible of a population in which both sexes interact with each other.

The finding of the ratio $r_{F/M} \simeq 2$ has also been seen as having important implications in relation to the difference in risk-taking behavior between genders. \citet{Baumeister} implicitly considered as equal the effective population size and the number of our ancestors. He made the general assumption that the two-to-one TMRCA ratio means that we have twice as many female as male ancestors. This interpretation implies important sex-biased behaviors, in particular in the area of risk taking. Baumeister argues that these biases help explain many puzzling facts about male social behavior, including competitive behaviors and excessive risk taking in modern financial markets. Another contribution of our ABM is to test Baumeister's interpretation concerning the number of our female and male ancestors.

Two classes of explanations have been proposed to explain the sex-biased genetic patterns that indicate unequal TMRCAs. One class of explanations is based on effective polygyny, `whereby males tend to father children with more females than females do with males' \citep{Hammer2004}. The other class is based on female sex-biased dispersal \citep{Seiel}.

For the first class, the most important consequence of effective polygyny is that males exhibit a higher variance in reproductive success than females. Indeed, as soon as some males monopolize several females each, others leave no descendants, while the distribution of reproductive success among females remains fairly egalitarian. Coalescent theory, which belongs to population genetics, established the theoretical material used to infer TMRCAs based on DNA samples (see \citet{Hedrick} for an introduction to coalescent theory as a part of population genetics). In coalescent theory, a larger male than female variance in reproductive success is expected to result in a smaller male than female TMRCA. Qualitatively, this is because some males have many more children than others, which makes the successful ones likely to become common ancestors. On the other hand, if all females have about the same number of children, it takes more time for a female to become a common ancestor (through what we could call `pedigree drift' in our model, by analogy with genetic drift). The two main population genetics models applied in the context of coalescent theiry are the Wright-Fisher model \citep{Wright1931, Fisher1930} and the Moran model \citep{Moran58, Moran62}. Each model assumes that individuals mate at random and predicts the TMRCA based on this plus other inputs (for a review of these two models, see \citet{Wakeley}).

The second class of explanations is female sex-biased geographic dispersal, whereby females disperse over larger spatial distances than males. This mechanism is often discussed as a possible explanation for observed patterns of genetic diversity in close non-human primates \citep{Erik2006} and in humans, some \citep{Seiel} arguing in favor of female dispersal, others \citep{Hammer2004, Hammer2004mig} dismissing it; see \citet{Wilkins} for a review. Migration of individuals results in a so-called gene flow, and coalescent theory including gene flow is called the `structured coalescent' \citep{Wakeley, Hedrick}. When females disperse, local mtDNA pools might include genes coming from a larger region, locally increasing genetic diversity. At the same time, if males do not migrate or migrate much less than females, NRY variation within populations should be quite small. Local DNA analyses then infer a higher female than male effective population size as more genetically different females than males contributed to each region, and thus deduce a higher female than male TMRCA. 

In the present paper, we follow a different reasoning and investigate how the mating structure influences the resulting TMRCA, using an agent-based model (ABM, or individual-based model, as it is often called in biology). Our agent-based model allows us to investigate which of several possible social organizations (including several mating systems and sex-biased migration) is/are most likely to result in larger female vs. male TMRCA. Our results focus on the TMRCA ratio $r_{F/M}=T_{F}/T_{M}$, which is about 2 according to genetic studies, as reviewed above. Our ABM is forward-looking in time and simulates a population with overlapping generations of agents. Our agents have life cycle events such as reaching sexual maturity or menopause. Our model includes no genetic evolution of individual traits. After each simulation lasting hundreds of generations, $T_{F}$ and $T_{M}$ are exactly retrieved by exploring the genealogical tree of the present male population backward in time, as if analyzing their mtDNA and NRY. We do not assume distributions of reproductive success from the start. Instead, we control the mating pattern at the social level and observe the resulting distributions of reproductive success for each gender. This is allowed by the unique features of our ABM, namely forward-looking and overlapping generations. This contrasts with coalescent models, which are backward-looking and assume non-overlapping generations. Existing forwards-in-time models almost always assume non-overlapping generations and thus have to input distributions of reproductive success. To our knowledge, the only program that implements overlapping generations is BottleSim and it has been applied exclusively to the problem of bottlenecks \citep{BottleSim}.

The different social organizations implemented in our ABM are polygynandry (or promiscuity) and polygyny (in the traditional sense, i.e., as a social contract and with some generalizations). Each mode comes with at least two variants, whereby the male population is respectively homogeneous or heterogeneous. In the latter case, some males are `alphas' and others are `betas', which can be understood as reflecting how easily they can mate based on female preferences. Input parameters are the proportion of alphas and how much more successful alphas are as compared to betas in attracting mates. A heterogeneous male population implies a competition among males as each of them strives to appear as an alpha. Effective polygyny (which by definition does not necessarily imply the existence of marriage in the form of a social contract) is or is not realized in our model, depending on the mode, variant and some agent life cycle characteristics. An additional variant in the polygyny mode allows to create a condition very close to lifelong monogamy. Furthermore, sex-biased migration can take place within any of these mating systems. We implement this possibility by considering an island model of two subpopulations (or demes) exchanging agents.

The exploration of the parameter space of our ABM shows that a TMRCA ratio $r_{F/M}$ of 2 is realized with a significantly higher likelihood within a heterogeneous male population (consisting of alpha and beta males) than within a homogeneous male population, independently of the mating system (polygynandry or polygyny). Indeed, while about 60\% of the simulations have $0.5<r_{F/M}<1.5$ in the homogeneous mode, a heterogeneous male population lowers by half this percentage and shifts the distribution of the TMRCA ratio to higher values. Moreover, the heterogeneity level needs to be high: less than half of the males can be alphas, while betas can have no more than half the fitness of alphas for the TMRCA ratio to depart from 1. Our simulations suggest that sex-biased migration alone is very unlikely to contribute to the observed TMRCA ratio. We also discuss the possibility that unequal sex-specific death rates may have played an important role, but argue that it could hardly have taken place with the required intensity without the existence of significant male-male competition.

The investigation of the genealogical tree produced by our ABM also shows that, in the modes that maximize the probability of the TMRCA ratio to lie between 1.5 and 2.5, the present generation has approximately 1.4 times as many female as male ancestors. An ancestor is defined as an agent who has at least one descendant among the present generation, along any lineage. Counting the ancestors belonging to each gender provides a measure of the transmission success of females and males, in a genealogical and potentially genetic sense. Our results support the interpretation of the observed TMRCA ratio made by \citet{Baumeister} and the sex-specific behavioral consequences thereof.

In summary, our results indicate that we are descended from males who succeeded in a highly competitive context and from females who did not have to face the same competitive conditions. We put these findings in the perspective of recent psychological and sociological studies concerning gender differences with respect to preference for competition, overconfidence and risk decisions. In particular, we link our results to sociological aspects of financial bubbles and crashes originating on trading floors that appear to be fertile grounds for unleashed male-male competition in the almost complete absence of females.

The remainder of this paper is organized as follows. We present our agent-based model in Section \ref{abm}. Detailed results are discussed in Section \ref{results}. Our main results are summarized in Section \ref{summary} and we discuss their implications in Section \ref{discussion}.

\section{Agent-based model} \label{abm}

\subsection{General set-up} \label{set-up}

\begin{figure*}
\begin{center}
 \includegraphics[width=.7\textwidth]{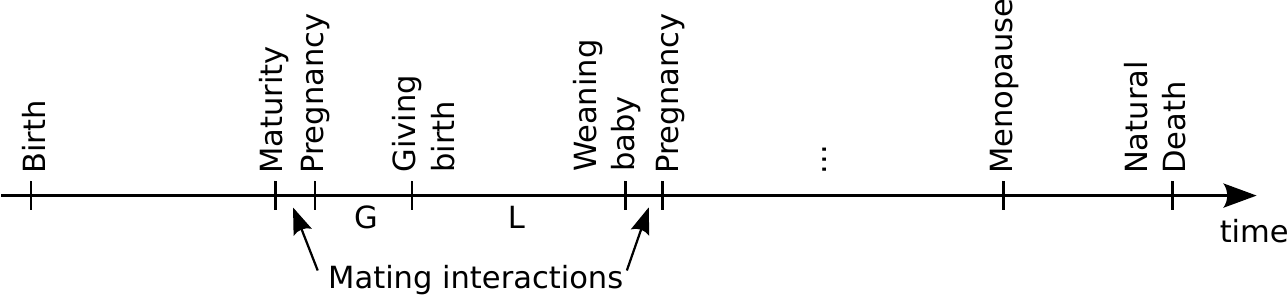}
\end{center}
\caption{Typical life cycle of a female (proportions are not respected). Males only have the events Birth, Maturity and Natural Death and participate to the mating interactions at every time step between Maturity and their death. When the carrying capacity is exceeded, any agent might die at any time between her birth and the age of natural death.}
\label{lifecycle}
\end{figure*}

Our agents have the following life cycles, as illustrated by Figure \ref{lifecycle}.

\textbf{Mating interactions} take place at every time step. Only mature males and mature premenopausal females participate. The different mating systems we implemented are described in section \ref{matingmodes}.

\textbf{Pregnancies and births (incl. lactation):} Pregnancy lasts G time steps. Pregnant females have the same probability of dying as any other agent. If this happens, the unborn baby is not counted as an agent in the simulation. Newborns have a 50\% chance of being of either sex. A mother remains infertile after giving birth during a lactation time L.

\textbf{Maturity and menopause:} When agents become sexually mature, they start participating in the mating interactions. Females stop when they undergo menopause.

\textbf{Deaths:} We give as inputs two carrying capacities $K_{F}$ and $K_{M}$, resp. for females and males. At each time step, we ensure that the population size of either sex is at most equal to the respective carrying capacity. The population thus remains constant with $K_{F}+K_{M}$ members in total. Agents that reach the maximum age die of natural death, independently of the population size. Specifically, if $N_{F}$ is the number of females (resp. males), the probability that a female survives, if she is younger than the maximum age, is equal to 1 if $N_{F} \leq K_{F}$ and to $K_{F}/N_{F}$ otherwise.

\textbf{Parameters:} Considering that one time step represents one calendar day, we choose $G=270$ ($\approx$ 9 months) and $L=1200$ ($\approx$ 3.5 years, which are typical of forager societies, e.g. see \citet{Konner}). Agents become sexually mature at 5'400 ($\approx$ 15 years old), females reach menopause at 18'000 ($\approx$ 50 years old) and natural death occurs at 25'000 ($\approx$ 70 years old). We checked that our results are not influenced by varying these parameters across realistic ranges, by having sex-specific ages of maturity or by letting menopausal females participate in the mating interactions (see section \ref{others}). We also discuss sex-specific carrying capacities ($K_{F} \neq K_{M}$) in section \ref{deathNbottleneck}. For most of our simulations, we set $K_{F}=K_{M}=75$, corresponding in total to Dunbar's prediction for the typical size of social groups associated with the human cognitive power to manage social relationships \citep{Dunbar}. As long as $K_{F}=K_{M}$, the population size  does not influence the results shown in Tables \ref{TableTMRCA} and \ref{TableRS} (we tested several population sizes up to $K_{F}=K_{M}=500$). No change of our results are found when reducing all duration-related parameters by a proportional factor, i.e. by setting that one time step represents for instance one week (then $G=38$, $L=170$, etc.) or one month ($G=9$, $L=40$, etc.). Thus, within a broad range, the granularity of time steps has no influence on our results.

We simulated hundreds of synthetic worlds for each set of population characteristics. For each simulation, we used the last generation of males as our sample and searched the genealogical tree for their female and male MRCA. As generations are overlapping, the generation length has to be externally defined in order to select our sample. This definition influences only our sample choice. We usually choose the generation length $l_{gen}$ equal to the maturity age and posit that agents born between iteration $n l_{gen}$ and $(n+1) l_{gen}$ belong to generation $n+1$ ($n \in \mathbb{N}$).

The number of time steps per simulation was chosen such that a female and a male common ancestors could always be found for the sample. With the above parameters, this corresponds to approximately 2 million time steps. Then, for each gender, the TMRCA was computed as the last time step minus the MRCA's date of birth. We observed that increasing the number of time steps does not increase TMRCAs but only shifts the MRCAs forward in time, as it should. 

In summary, our parameters are as follows:
\begin{eqnarray} \label{parameters}
&& \textrm{1 time step} \simeq \textrm{1 day} ; \nonumber\\
&& G=270 ; L=1200 ; \textrm{Menopause}=18000 ; \nonumber \\
&& \textrm{Maximum Age}=25000 ; \nonumber \\
&& K_{F}=75 ; K_{M}=75 ; \textrm{Iterations}=2 \times 10^{6}
\label{eq:sans_multline}
\end{eqnarray}

\subsection{Male population characteristics}

Each newborn male is assigned a fitness $f \in \left[0,1\right]$. It can represent his technical skills (such as hunting or buiding a shelter), social status, attractiveness, physical strength/endurance, sociability, health, intelligence, resources or many other traits. Depending on the mating mode, this fitness acts as the probability that fertile females met by a male get pregnant from him, or as the probability that females choose a male to be their husband.

Alpha males have a fitness $f_{\alpha}=1$ and beta males have a fitness $f_{\beta}$ with $0<f_{\beta}<1$ (all beta males have the same fitness). Newborn males have a probability $p_{\alpha}$ of being alphas and $1-p_{\alpha}$ of being betas, independently of their father's fitness.

If $p_{\alpha}=1$ or $p_{\alpha}=0$, the male population is homogeneous (all-alphas or all-betas). Otherwise, we refer to this situation as a heterogeneous mode characterized by $p_{\alpha}$ and $f_{\beta}$. 

\subsection{Mating interactions and sex-biased migration} \label{matingmodes}

We tested our model with two different mating systems, polygynandry and polygyny, as well as under a condition of sex-biased migration.

\subsubsection{Polygynandry} \label{polygynandry}

At each time step, each mature male chooses a fertile female at random and tries to mate. This results in a pregnancy with a probability equal to the male's fitness. The order in which males are considered to pick a female is randomized at each time step. It is thus as if males were choosing females at the same time during each time step and sometimes competiting over the same female, with the winner being chosen according to his fitness, tights being resolved by random selection.

This mating pattern can be associated with polygynandry (or promiscuity), as there is no monitoring of how many sexual partners an agent can have in his or her lifetime. In a homogeneous male population, it is equivalent to fully random mate choice.

\subsubsection{Polygyny} \label{SecPolygyny}

In the polygyny mode, a female who becomes mature chooses one of the mature males to be her husband, who will then father all of her children for as long as he lives. Females pick the male $i$ who can offer them the largest `adjusted fitness' $F_{i}$, defined as
\begin{equation}
F_{i} := \frac{f_{i}}{w_{i}+1},
\end{equation}
where $f_{i}$ is the male's fitness and ${w_{i}}$ counts how many wives he already has. If several males have the maximum adjusted fitness, females randomly choose one of them. This follows the concept of polygyny threshold \citep{PolThresh66, PolThresh69} whereby a female might gain from marrying a male who already has one or several wives if the amount of resources he has to offer offsets the disadvantage of having to share with other wives, in comparison to choosing a single male with less resources.

This mode comes together with two variants. First, if one of a male's wives dies, she remains counted or not in $F_{i}$. Still counting a deceased wife in $F_{i}$ can be interpreted as taking into account that she most likely left her husband with at least one child to take care of, decreasing the amount of resources (encapsulated in $f_{i}$) he can offer to a new wife. This option will be abbreviated DWC (Deceased Wives Count) versus DWnoC (Deceased Wives Don't Count). 

Second, we can allow fertile females to remarry or not when their husband dies, in which case female choices are performed on the same basis as above. We will denote this option by WREM (Widows Remarry) or WnoREM (Widows Don't Remarry). Menopausal females whose husband dies never remarry, whatever the variant. WnoREM can be interpreted as integrating either a lack of male interest in widows that most likely already have children from their first husband, or a lack of female interest in remarrying.

In contrast to the polygynandry mode, polygyny integrates explicit pair bonds, whereby one male can be paired simultaneously to several females and females are always paired to one single male at a time. Actually, one of the polygyny variants turns out to be very close to lifelong monogamy, as shown in section \ref{resPolygyny}. Anthropological studies indicate that the prevalent mating systems in humans are polygyny and monogamy \citep{Marlowe}, which account for approximatively 82\% and 17\% of human societies, respectively. In addition, pair bonding and its counterpart, cooperative breeding, are essential features of humans \citep{vanSchaik}. Therefore, our polygyny mode simulating pair bonds seems more realistic than polygynandry. However, by its simplicity, the polygynandry mode remains a useful testing tool. 

\subsubsection{Sex-biased migration}

We tested a two-deme island model with sex-specific migration rates $m_{F}$ for females and $m_{M}$ for males. An agent can migrate only once in its lifetime, just before it reaches sexual maturity. The probability that it migrates at this time is given by its gender's migration rate.

\section{Results} \label{results}

With the parameters discussed in section \ref{abm}, the total number of agents born during a typical simulation is about 44'000. If a generation is set as the generation of the father plus 1 for a boy and as the generation of the mother plus 1 for a girl (the initial agents belonging to generation 0), 2 million time steps give about 200 generations with approximately 100 agents of each sex belonging to each generation. We measured inbreeding defined as the proportion of matings taking place between siblings or parent-child. In all modes, it is found below 6\%.

\subsection{Polygynandry}

\begin{figure*}
\begin{center}
 \includegraphics[width=.9\textwidth]{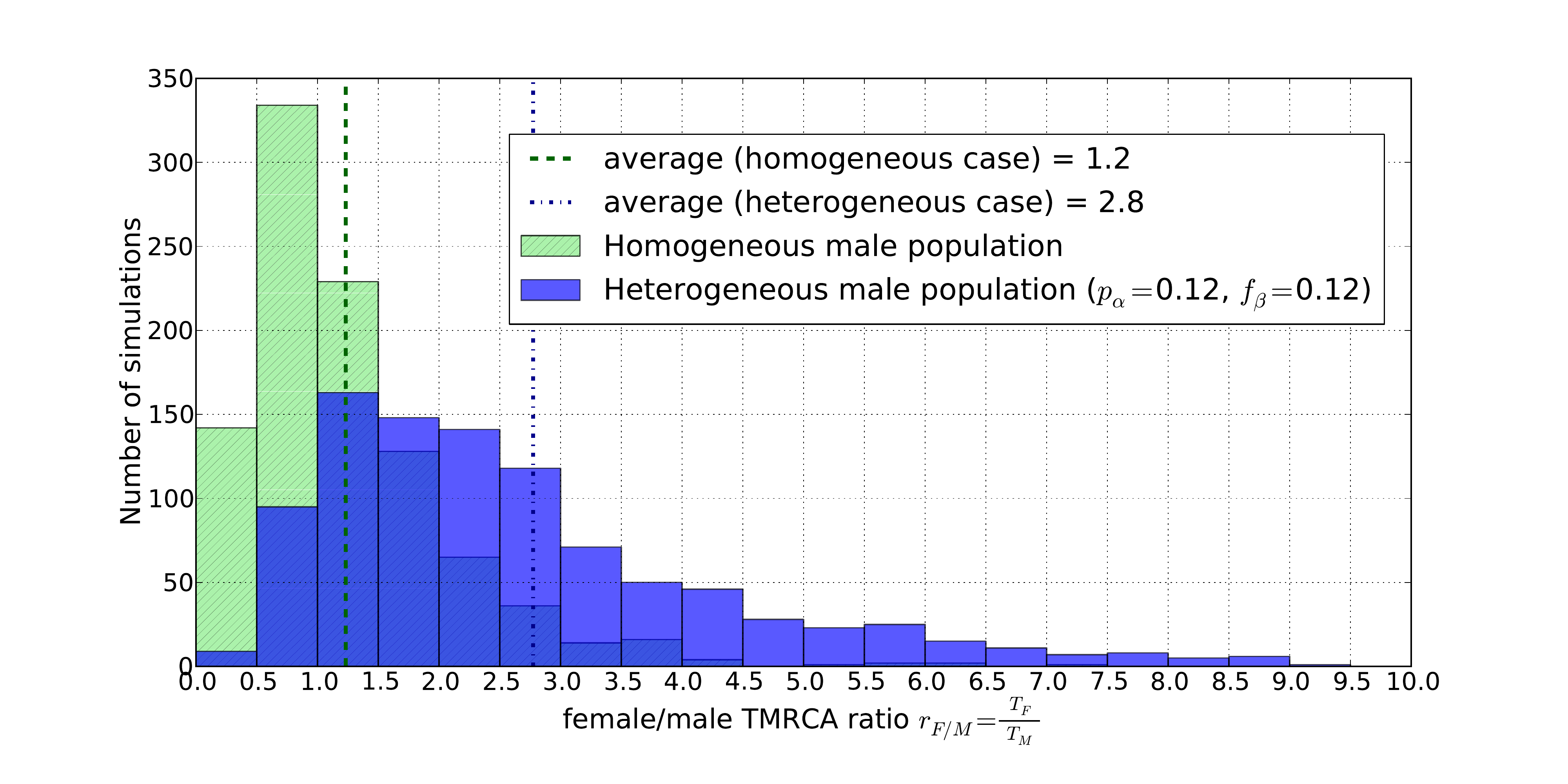}
\end{center}
\caption{Histogram of the TMRCA ratio $r_{F/M}=T_{F}/T_{M}$ for a polygynandrous mating system with a homogeneous male population (hatched area) and with a heterogeneous male population when $p_{\alpha}=0.12, f_{\beta}=0.12$ (plain area). The average of each distribution is indicated by dotted lines. Each histogram is obtained over 1000 simulations. Parameters are given in Expression \ref{parameters}.}
\label{Polygynandry_homo_1000runs}
\end{figure*}

Among one thousand simulations of a polygynandrous mating system in a homogeneous male population ($p_{\alpha}=1$) and equal carrying capacities (so that death rates are equal for males and females), we find 58\% of them with a TMRCA ratio $r_{F/M}=T_{F}/T_{M}$ between 0.5 and 1.5, compared with 20\% between 1.5 and 2.5 (Figure \ref{Polygynandry_homo_1000runs} and Table \ref{TableTMRCA}). The distribution of the TMRCA ratio shows a mean of 1.2 and a median of 1, with a standard deviation of 0.8 reflecting the high stochasticity of the process.

\begin{figure}
\begin{center}
 \includegraphics[width=.5\textwidth]{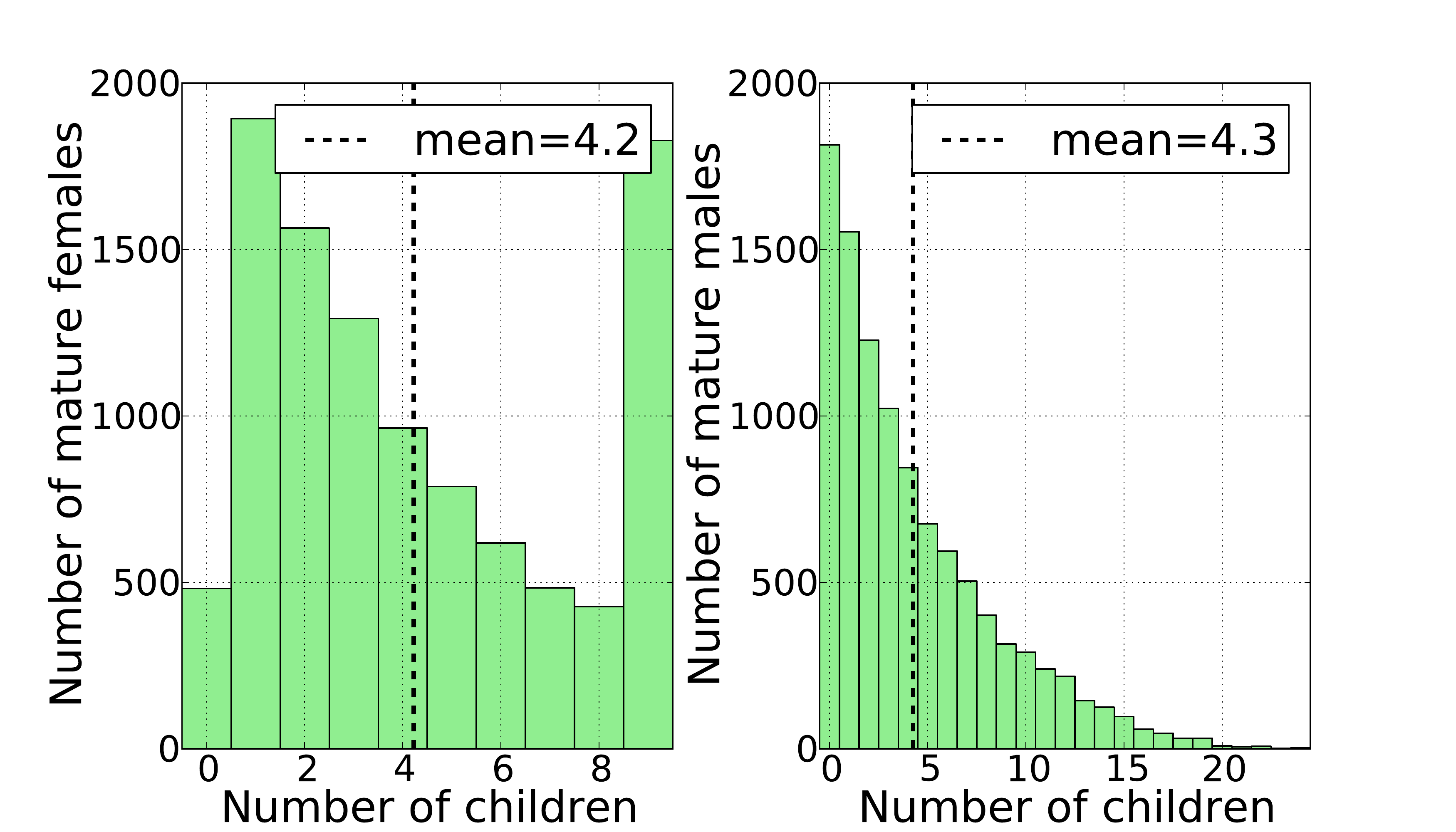}
\end{center}
\caption{Histogram of the number of children per mature female (left panel) and per mature male (right panel) in the polygynandry case with a homogeneous male population, for one specific simulation. As the population is kept constant, the average number of children per female and per male is 2, but here only the mature agents' numbers of children are shown, which is why the averages are around 4. The maximum number of children a female can have between maturity and menopause is 9, and all females who reach menopause have 9 children. 5\% of mature females are childless because they die within G time steps after reaching maturity. In comparison, 18\% of the mature males die before they can have any children. Parameters are given in Expression \ref{parameters}.}
\label{Polygynandry_homo_Children}
\end{figure}

We use the number of children born to an agent in her lifetime as a measure of her reproductive success. We consider only agents that reached maturity, as agents that died while still immature could not have any children. As the population is kept constant, the average number of children per agent of either sex is 2, but the average number of children per mature agent is approximately 4 (see Table \ref{TableRS}). In the polygynandry mode with homogeneous male population (corresponding to random mating), males have a larger variance in reproductive success than females\footnote{We use the term `variance' because of its common use in the context of measuring reproductive success. However, it should be kept in mind that the variance of a random variable is the square of its standard deviation, and it is the standard deviation that should be used as the measure of variability that can be compared meaningfully with the typical (or average) value of the random variable.}. Indeed, as shown in Table \ref{TableRS}, the standard deviation in reproductive success is on average 4.2 for mature males, compared with 3 for mature females (the corresponding variances for males and for females are respectively 17.7 and 8.8). This gender difference takes place despite random mating because female reproductive rate is severely limited in time by biological constraints (namely pregnancy and lactation), while male reproductive rate is only limited by the number of mates a male can access in his lifetime. In other words, female have a longer mating time than males, as in Sutherland's review of Bateman's data \citep{Sutherland,Bateman}. Yet, as mature females are most of the time unavailable (pregnant, lactating or menopausal), even the most successful male of each simulation cannot have, on average, more than 26 children. Figure \ref{Polygynandry_homo_Children} shows the number of children born to mature females and males for one specific simulation.

\begin{figure}
\begin{center}
 \includegraphics[width=.5\textwidth]{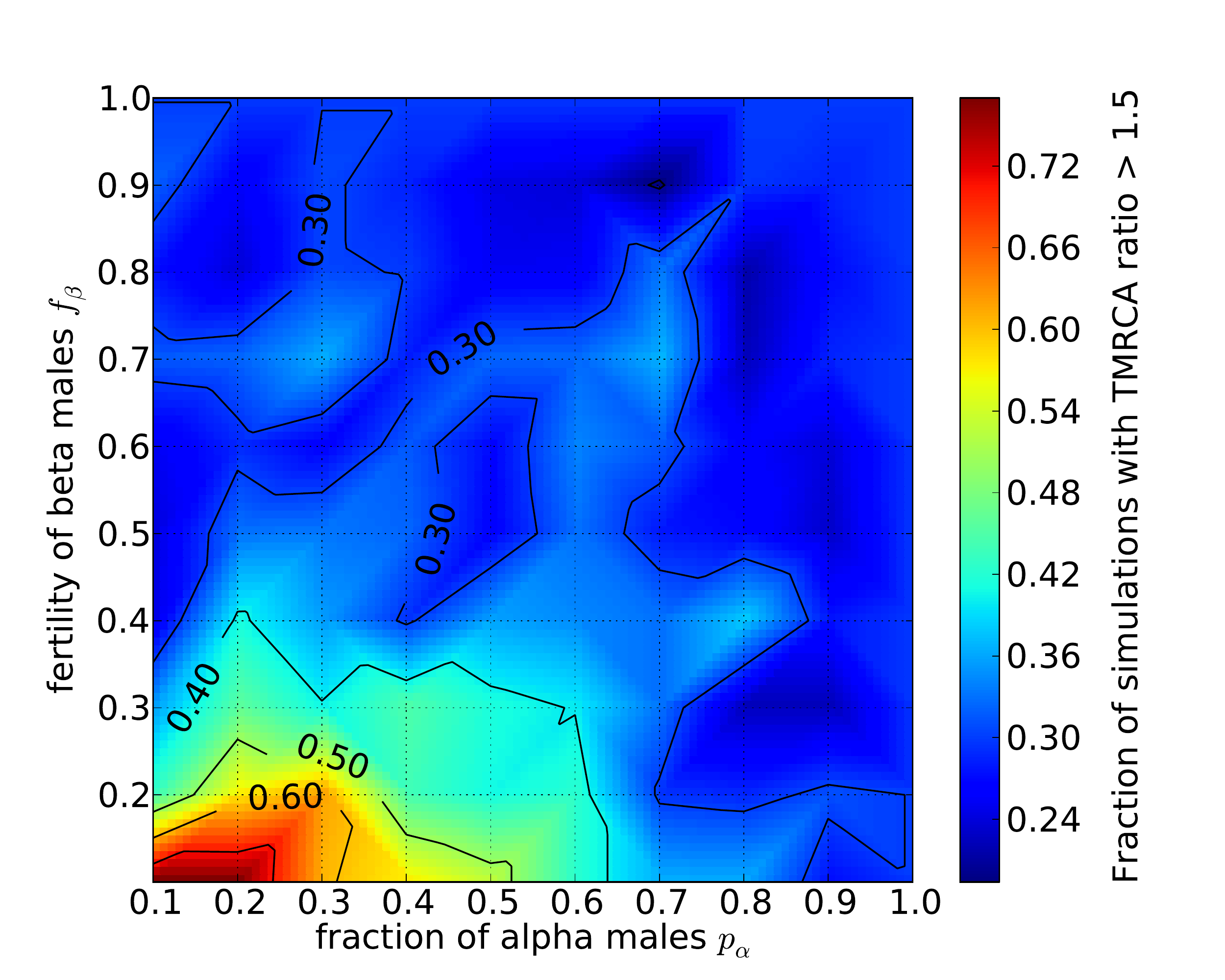}
\end{center}
\caption{Phase diagram in the case of polygynandry with heterogeneous male population, giving the fraction of simulations with a TMRCA ratio $r_{F/M}=T_{F}/T_{M}$ larger than 1.5 in the two-parameter space (fraction of alpha males $p_{\alpha}$; fertility of beta males $f_{\beta}$). Each grid point is obtained over 200 simulations. Parameters are given in Expression \ref{parameters}.}
\label{griddata_polygynandry_larger}
\end{figure}

\begin{figure}
\begin{center}
 \includegraphics[width=.5\textwidth]{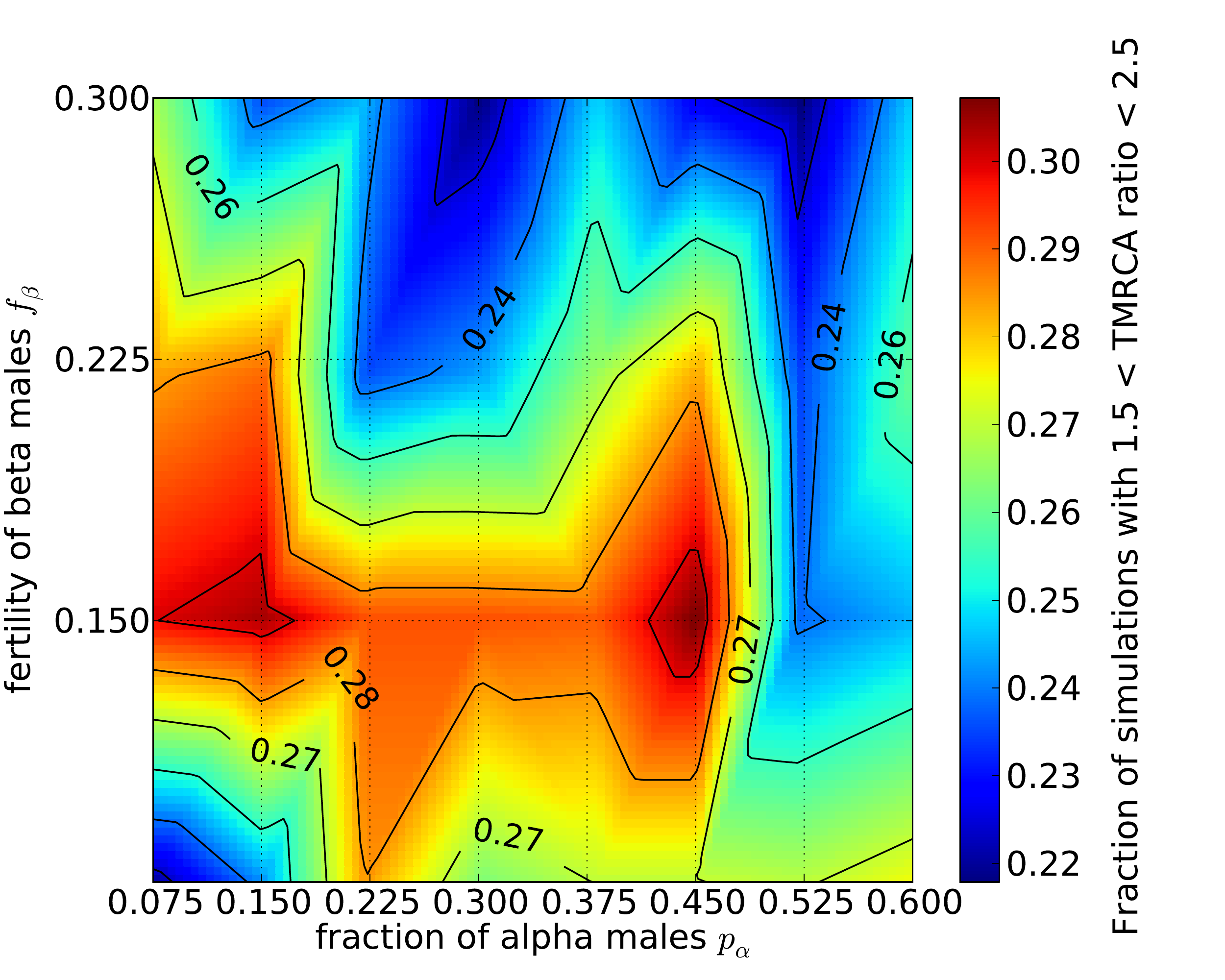}
\end{center}
\caption{Phase diagram in the case of polygynandry with heterogeneous male population, giving the fraction of simulations with a TMRCA ratio $r_{F/M}=T_{F}/T_{M}$ between 1.5 and 2.5 in a region of the two-parameter space (fraction of alpha males $p_{\alpha}$; fertility of beta males $f_{\beta}$). Each grid point is obtained over 500 simulations. Parameters are given in Expression \ref{parameters}.}
\label{griddata_polygynandry_zoom_between}
\end{figure}

Polygynandry in a heterogeneous male population gives a very different distribution of TMRCA ratios. Recall that male population heterogeneity is characterized by the two parameters $p_{\alpha}$ and $f_{\beta}$. Figure \ref{Polygynandry_homo_1000runs} shows a superposition of the distribution of the TMRCA ratio in the homogeneous case and in the heterogeneous case with $p_{\alpha}=0.12$ and $f_{\beta}=0.12$. Figures \ref{griddata_polygynandry_larger} and \ref{griddata_polygynandry_zoom_between} show that, for these values, it is possible to obtain more than 72\% of the simulations with a TMRCA ratio higher than 1.5 and up to 30\% of the simulations with a TMRCA ratio between 1.5 and 2.5, if $p_{\alpha} \lessapprox 0.5$ and $f_{\beta} \lessapprox 0.3$. 

In Table \ref{TableRS}, we report some characteristics of the male and female reproductive successes when $p_{\alpha}=0.12$ and $f_{\beta}=0.12$. In comparison to a homogeneous male population, the variance in reproductive success among mature males explodes: the standard deviation is now equal to 8 and the most successful males have on average 82 children, while the situation has not changed for females. Table \ref{TableTMRCA} gives some characteristics of the distribution of the TMRCA ratio when $p_{\alpha}=0.12$ and $f_{\beta}=0.12$. The average TMRCA ratio is now 2.8 (median 2.2) and only 26\% of the simulations show a TMRCA ratio between 0.5 and 1.5, compared with 29\% between 1.5 and 2.5.

\subsection{Polygyny} \label{resPolygyny}

Polygyny can be implemented in a homogeneous or heterogeneous male population and with four possible variants. With a homogeneous male population, the following behaviors are observed.

\begin{figure}
\begin{center}
 \includegraphics[width=.5\textwidth]{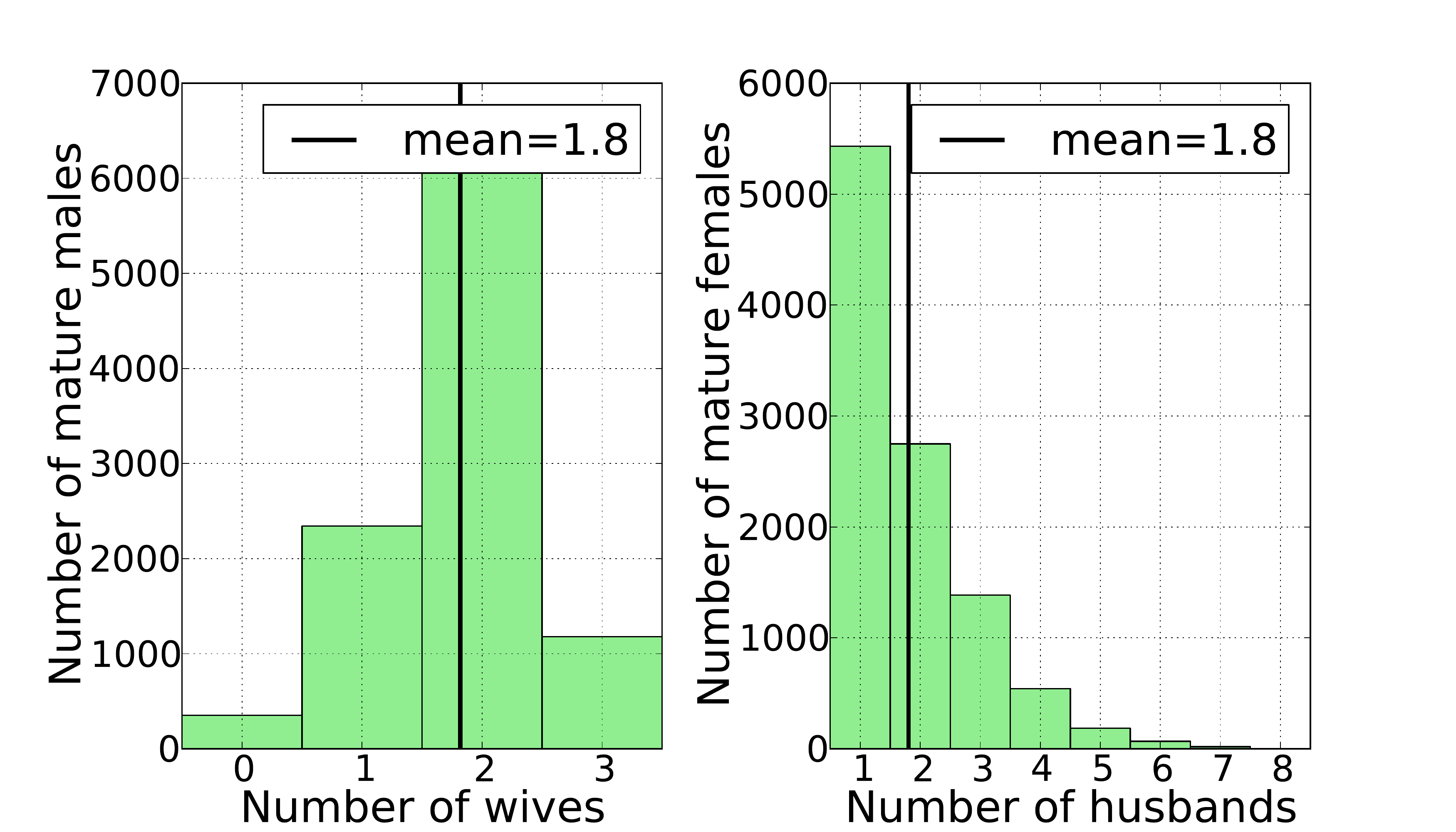}
\end{center}
\caption{Histogram of the number of spouses per agent in the polygyny mode with homogeneous male population, variant DWC/WREM (deceased wives decrease a male's adjusted fitness; with widows remarriage). Left panel: number of wives per mature male, right panel: number of husbands per mature female. In comparison to the situation where widows cannot remarry (DWC/WnoREM), males get more wives because there are more females looking for a husband (namely the newly mature females plus the widows). Parameters are given in Expression \ref{parameters}.}
\label{Polygyny_homo_WREM_Spouses}
\end{figure}

If deceased wives decrease a male's adjusted fitness and widows remarry (DWC/WREM), this variant is equivalent to sequential polyandry together with mild polygyny. Indeed, because of random female choices and slightly fluctuating population sizes of mature females and males, some males have several wives at the same time and females have several husbands in their lifetime (but always one at a time). However, at some point a male might not be chosen as a husband anymore because he has too many deceased wives, while a fertile female will never be without a husband. This variant gives the highest average TMRCA ratio $\bar{r}_{F/M}=1.5$, a median of 1.2, 52\% of the simulations with $0.5<r_{F/M}<1.5$ and 24\% of the simulations with $1.5<r_{F/M}<2.5$ (Table \ref{TableTMRCA}). Figure \ref{Polygyny_homo_WREM_Spouses} shows the distributions of the number of spouses per agent over the agents' lifetime.

\begin{figure}
\begin{center}
 \includegraphics[width=.5\textwidth]{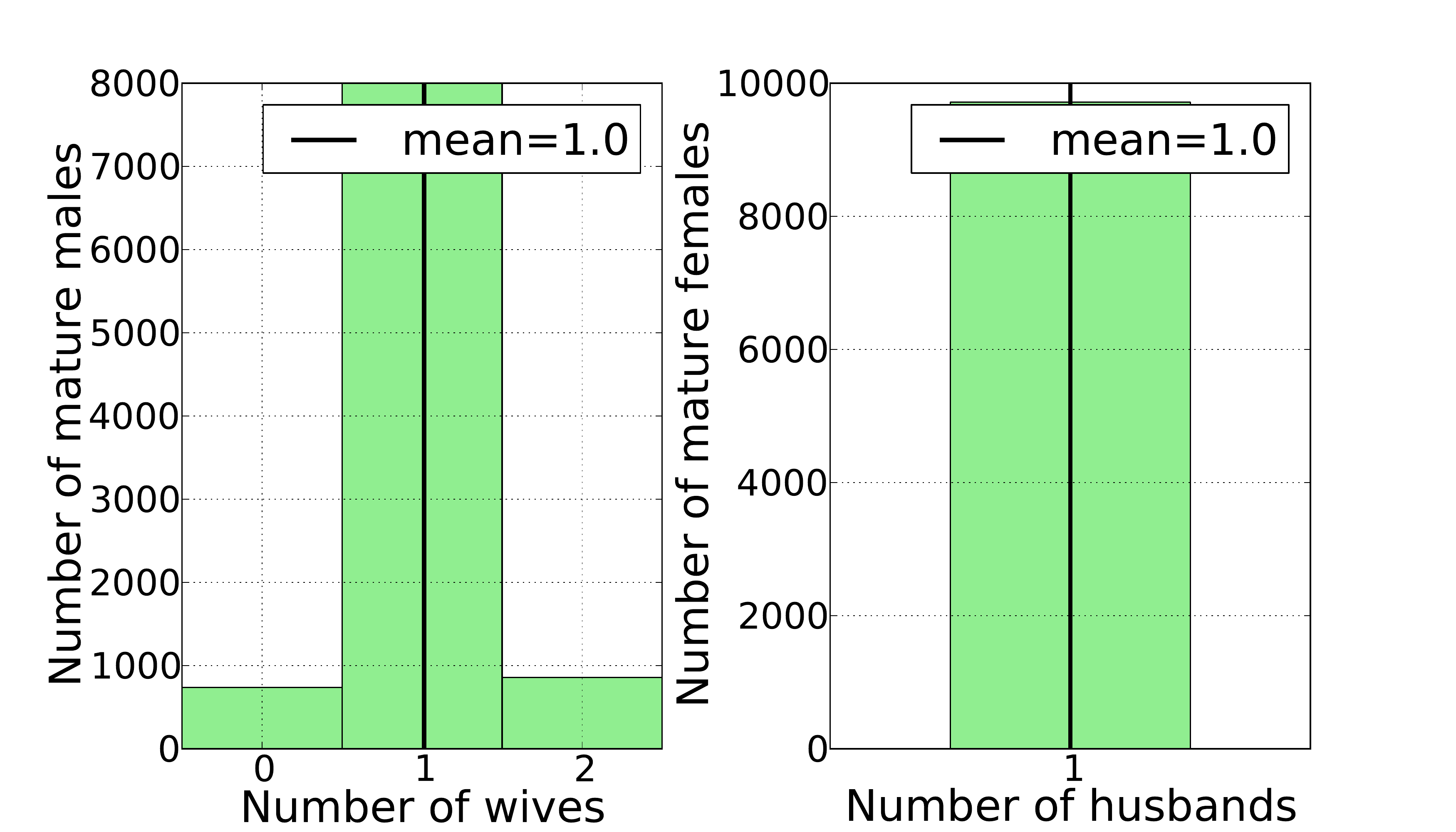}
\end{center}
\caption{Histogram of the number of spouses per agent in the polygyny mode with homogeneous male population, variant DWC/WnoREM (deceased wives decrease a male's adjusted fitness; no widows remarriage). Left panel: number of wives per mature male, right panel: number of husbands per mature female. Males do not get many wives because females do not seek remarriage if their husband dies. The only females looking for a husband are the newly mature ones, and since there are about as many mature females as males, this mode is very close to lifelong monogamy. Parameters are given in Expression \ref{parameters}.}
\label{Polygyny_homo_WnoREM_Spouses}
\end{figure}

If females are not allowed to remarry (DWC/WnoREM), polygyny is very weak because males are little solicited, there is no polyandry whatsoever and this variant becomes very close to lifelong monogamy (Figure \ref{Polygyny_homo_WnoREM_Spouses}).

If deceased wives do not influence the males' adjusted fitnesses and if widows remarry (DWnoC/WREM), males can have more wives than in the first variant, so we can expect sequential polyandry and medium polygyny.

These two last variants have in common that they both make genders as equivalent as they can get in the polygyny mode, DWC/WnoREM being close to monogamy, and DWnoC/WREM allowing everyone to remarry without being penalized by past marriages. This is why the average TMRCA ratio is close to 1 for both variants (Table \ref{TableTMRCA}). 

Finally, in the DWnoC/WnoREM variant, females have only one opportunity to marry, while males not only can have several wives but are not penalized by their late wives to attract new ones. With this variant, a male can find it easier to become a common ancestor than a female, which is reflected by the fact that the average TMRCA ratio $r_{F/M}$ is smaller than 1 (Table \ref{TableTMRCA}). 

In summary, polygyny in a homogeneous male population, whatever the variant, gives an average TMRCA ratio smaller than those obtained with heterogeneous male populations in polygynandrous societies (see Table \ref{TableTMRCA}). Similarly to polygynandry in a homogeneous male population, the distributions of the number of children per mature female and male suggest that the variance in reproductive success of males, although higher than that of females (Table \ref{TableRS}), is too much limited by females' unavailability to allow the TMRCA ratio to take off. 

\begin{figure}
\begin{center}
 \includegraphics[width=.5\textwidth]{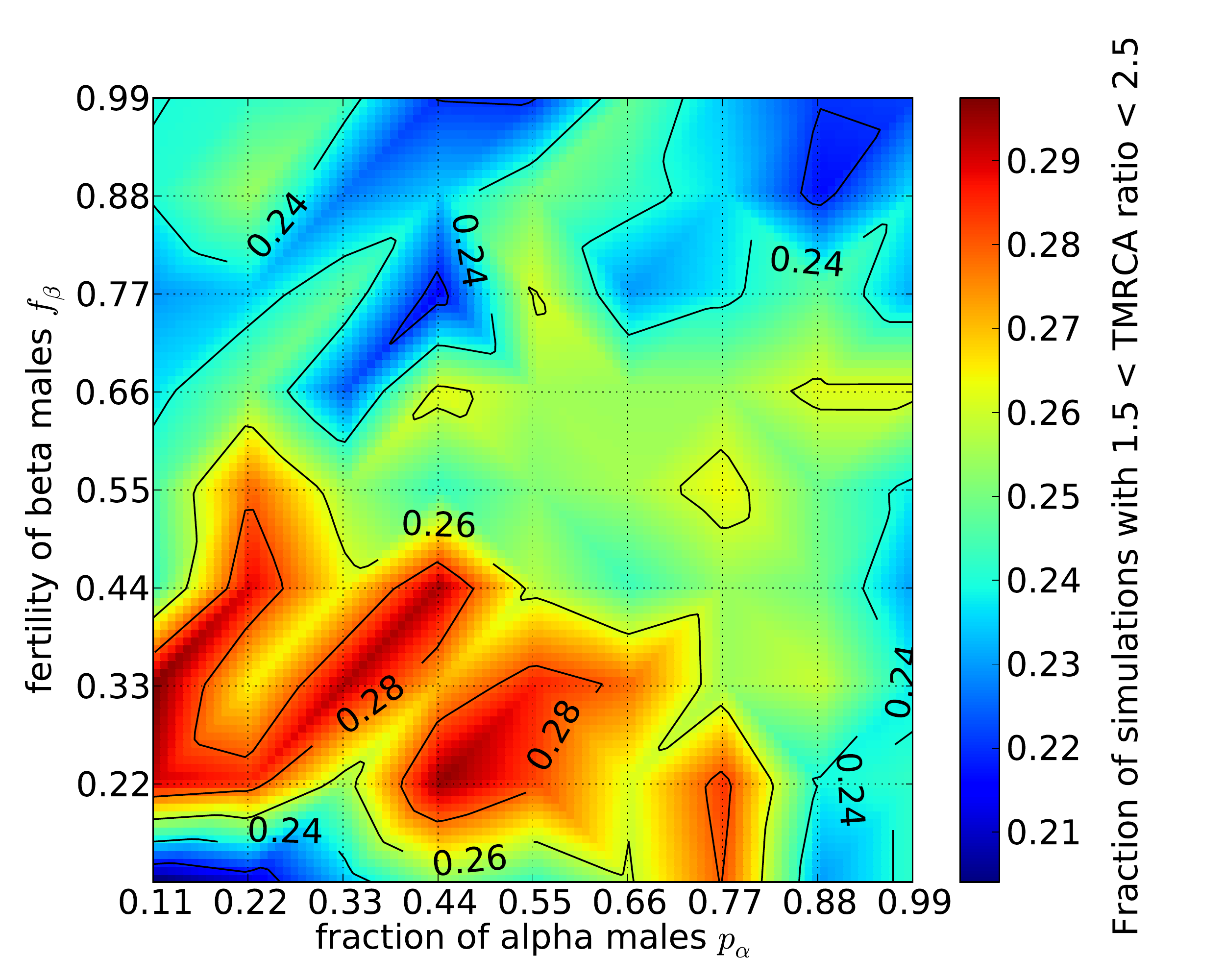}
\end{center}
\caption{Phase diagram in the case of polygyny with heterogeneous male population, deceased wives decreasing a male's adjusted fitness and widows remarriage (DWC/WREM), giving the fraction of simulations with a TMRCA ratio $r_{F/M}=T_{F}/T_{M}$ between 1.5 and 2.5 in the two-parameter space (fraction of alpha males $p_{\alpha}$; fertility of beta males $f_{\beta}$). Each grid point is obtained over 1000 simulations. Parameters are given in Expression \ref{parameters}.}
\label{images/griddata_Polygyny_DWC_WREM_between}
\end{figure}

\begin{figure}
\begin{center}
 \includegraphics[width=.5\textwidth]{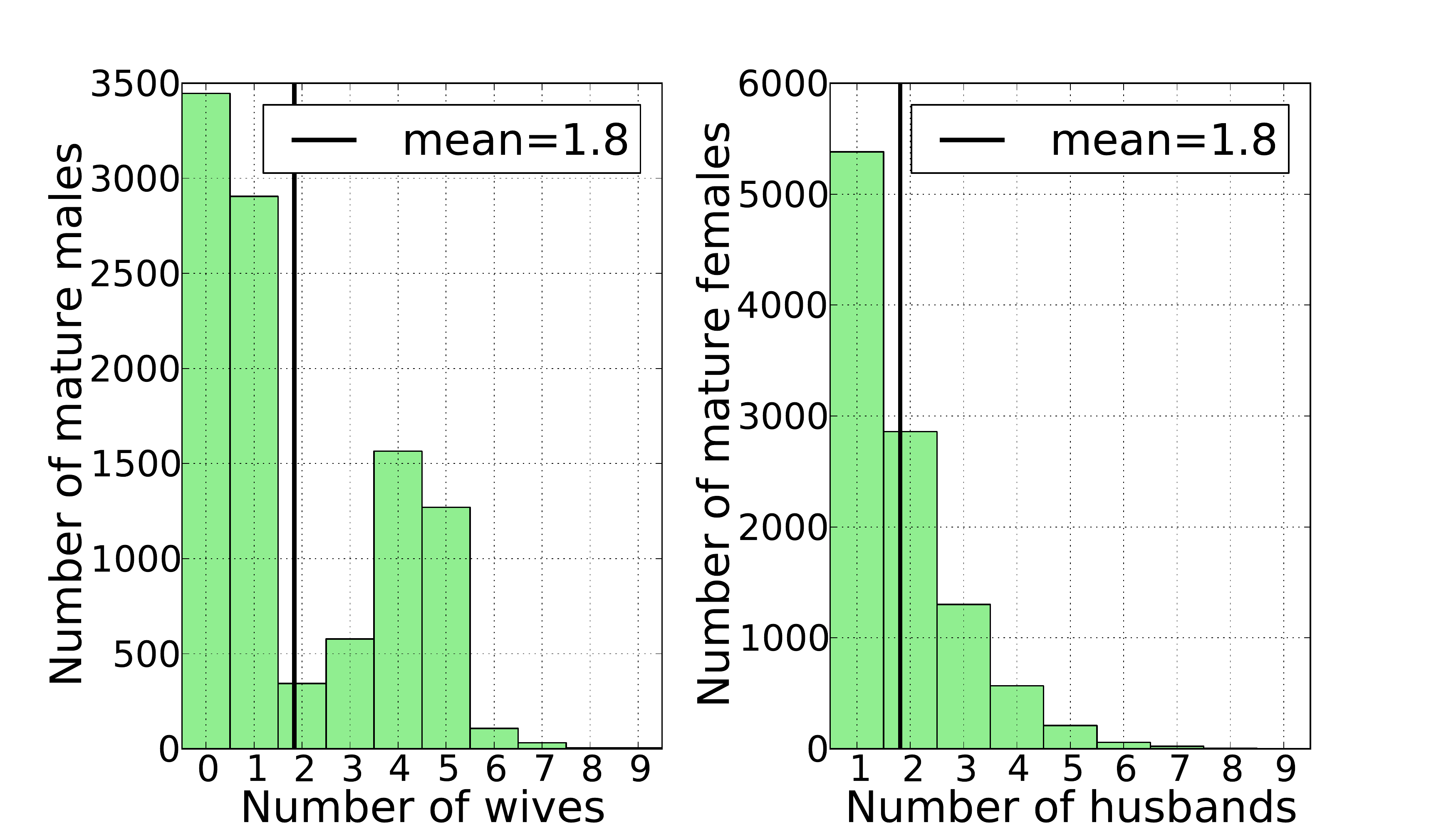}
\end{center}
\caption{Histogram of the number of spouses per agent in the polygyny mode with heterogeneous male population ($p_{\alpha}=0.4$, $f_{\beta}=0.2$), variant DWC/WREM (deceased wives decrease a male's adjusted fitness; with widows remarriage). Left panel: number of wives per mature male, right panel: number of husbands per mature female. Compared with the same variant in a homogeneous male population, the situation for females has not changed, while alpha males can now have up to 9 wives. There are many unmarried males because for each male who has $n$ wives, there are $n-1$ males with no wife (if there are as many mature males as females). The bimodal distribution of the number of wives per male reveals the two populations of alpha and beta males. Parameters are given in Expression \ref{parameters}.}
\label{Polygyny_compet_WREM_Spouses}
\end{figure}

To investigate the impact of a heterogeneous male population, we choose the variant DWC/WREM as it gives the highest average TMRCA ratio with a homogeneous population. This variant with heterogeneous male population is strongly polygynous, as alpha males attract many more wives than beta males. The TMRCA ratio is now between 1.5 and 2.5 in up to 30\% of the simulations, as long as $p_{\alpha} \lessapprox 0.5$ and $f_{\beta} \lessapprox 0.5$ (Figure \ref{images/griddata_Polygyny_DWC_WREM_between}). Tables \ref{TableTMRCA} and \ref{TableRS} show characteristics of the distribution of the TMRCA ratio and the numbers of children per mature female and male for $p_{\alpha}=0.4$, $f_{\beta}=0.2$. The average TMRCA ratio is 2.9 and the median is 2.3. As in the polygynandry mode with heterogeneous male population, 26\% of the simulations have a TMRCA ratio between 0.5 and 1.5, compared with 29\% between 1.5 and 2.5. Figure \ref{Polygyny_compet_WREM_Spouses} shows the distributions of the number of spouses per female and male for $p_{\alpha}=0.4$ and $f_{\beta}=0.2$.

\subsection{Sex-biased migration}

\begin{figure}
\begin{center}
 \includegraphics[width=.5\textwidth]{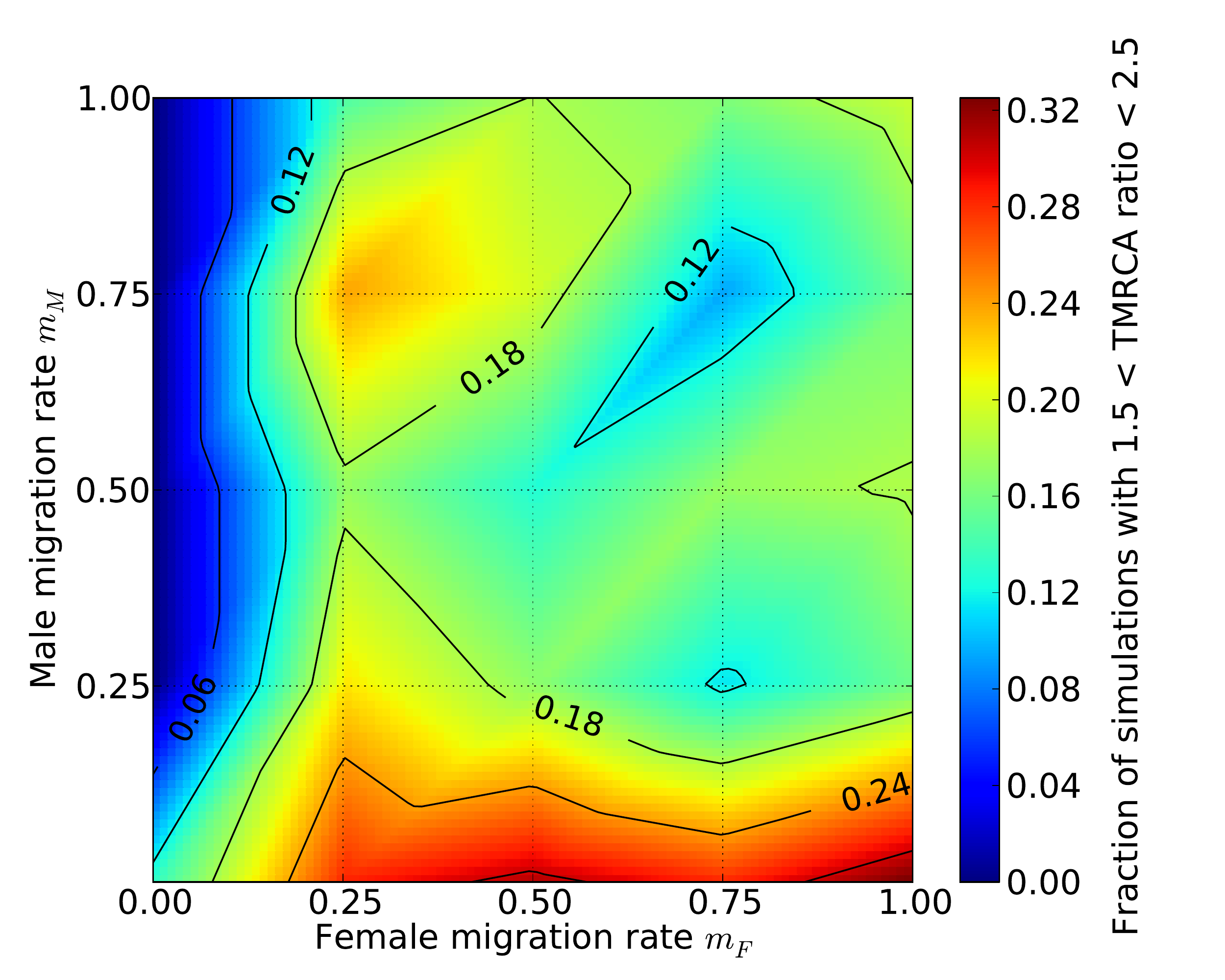}
\end{center}
\caption{Phase diagram in the case of polygynandry with homogeneous male population and migration, giving the fraction of simulations for one deme with a TMRCA ratio $r_{F/M}=T_{F}/T_{M}$ between 1.5 and 2.5 in the two-parameter space (female migration rate $m_{F}$; male migration rate $m_{M}$). Each grid point is obtained over 200 simulations. Parameters are given in Expression \ref{parameters}.}
\label{griddata_Migration_between}
\end{figure}

We tested sex-biased migration in a two-deme island model of polygynandrous societies with homogeneous male populations. As shown in Figure \ref{griddata_Migration_between} which represent the results for one of the two demes chosen arbitrarily, about 30\% of the simulations have a TMRCA ratio between 1.5 and 2.5 if $m_{F}>0.1$ and $m_{M}=0$, where $m_{F}$ and $m_{M}$ are respectively the female and male migration rates. Table \ref{TableTMRCA} presents the main properties of the distribution of the TMRCA ratio for one of the two demes when males never migrate and 10\% of the females do. The average TMRCA ratio is 2.1 with a median of 1.7. However, there are still more simulations with ratios between 0.5 and 1.5 than between 1.5 and 2.5 (37\% vs. 32\%).

A non-zero female migration rate together with a zero male migration rate means that `as soon as' (backward in time) one branch of the backward all-females ancestry lines exits deme A to enter deme B, we can find a female common ancestor for deme A only if either this branch comes back into A, or all-females ancestry lines from A all enter B and coalesce there. This is why female dispersal in the absence of male dispersal requires more time (backwards) to find a female common ancestor.

In the blue region of Figure \ref{griddata_Migration_between} where $m_{F}=0$ and $m_{M}>0$, the TMRCA ratio is not really zero but in fact does not exist because no MRCA female could ever be found for our sample (the last generation of males). Indeed, if any male from the sample of one of the demes (say A) was born in the other deme (B) and migrated later to A, his mother necessarily belonged to deme B. Since the female migration rate is zero, this female's all-females ancestry line is entirely in B and will never coalesce with the all-females ancestry lines of the other males from the sample in A, which are entirely in A.

\subsection{Sex-specific death rates and sex-biased bottleneck} \label{deathNbottleneck}

\begin{figure*}
\begin{center}
 \includegraphics[width=.9 \textwidth]{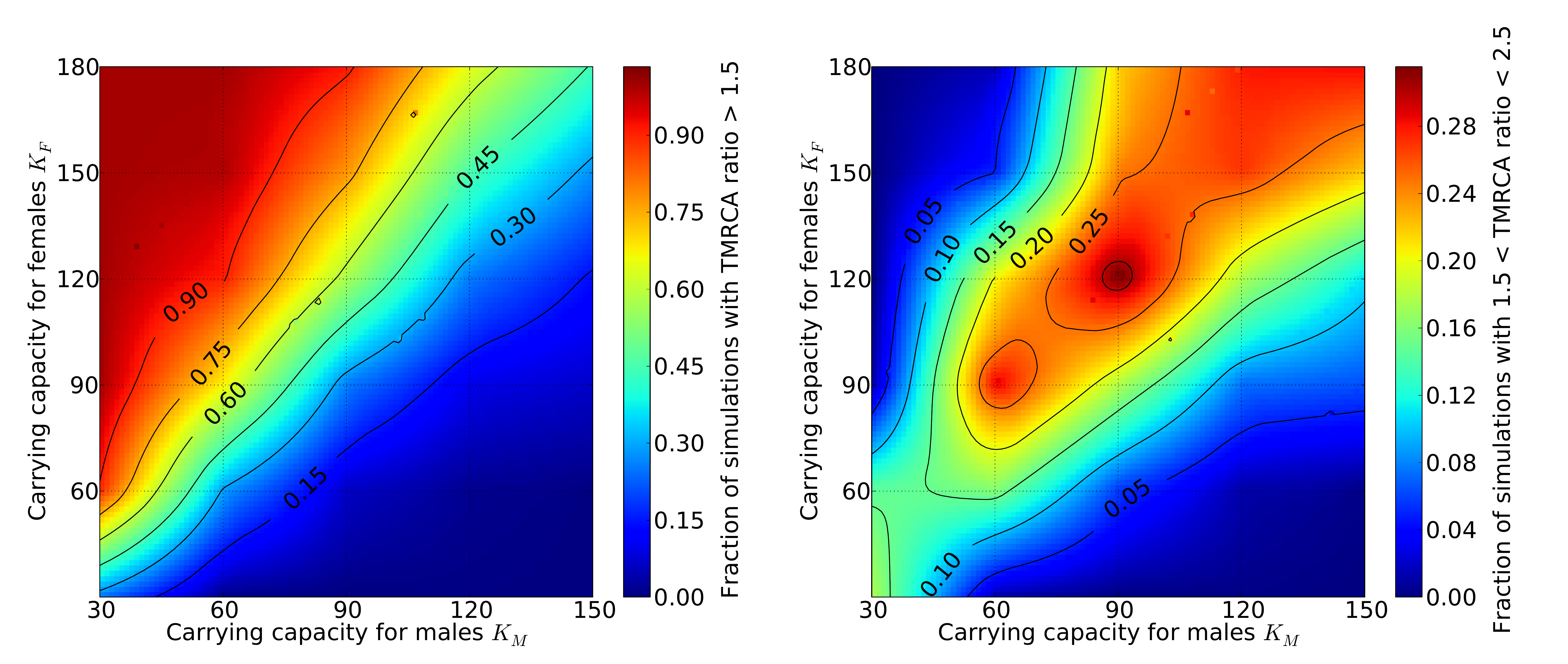}
\end{center}
\caption{Phase diagram in the case of polygynandry with homogeneous male population and sex-specific carrying capacities, giving the fraction of simulations with a TMRCA ratio $r_{F/M}=T_{F}/T_{M}$ larger than 1.5 (left panel) and between 1.5 and 2.5 (right panel) in the two-parameter space (carrying capacity for males $K_{M}$; carrying capacity for females $K_{F}$). Each grid point is obtained over 500 simulations. Parameters are given in Expression \ref{parameters}, except for $K_{M}$ and $K_{F}$ that are varied as indicated in the phase diagrams.}
\label{griddata_KfKm_2}
\end{figure*}

There is another way than sex-biased migration to shift the distribution of the TMRCA ratio $r_{F/M}=T_{F}/T_{M}$ to higher values in a homogeneous male population (with any mating system), which is to set a smaller male than female carrying capacity, so that there is a higher male than female death rate (the sex of newborns being randomly chosen with equal probability). Figure \ref{griddata_KfKm_2} shows that about 30\% of the simulations have a TMRCA ratio between 1.5 and 2.5 when $K_{F} \approx 1.5 \cdot K_{M}$. For example, with $K_{F}=120$ and $K_{M}=75$, we get 23\% of the simulations with $0.5<r_{F/M}<1.5$, 30\% with $1.5<r_{F/M}<2.5$ and 45\% of the simulations above 2.5 (see Tables \ref{TableTMRCA} and \ref{TableRS}).

Death causes reported in studies of hunter-gatherer societies are usually categorized into four classes: diseases (illness or parasites), degenerative or congenital diseases (including childbirth and old age), accidents and violence \citep{HiwiMortality, AcheMortality}. The main sex difference observed in a mortality study among the Hiwi hunter-gatherers of Venezuela \citep{HiwiMortality} is manifest for people older than 40 years old and is mainly due to violence (19 deaths per 1000 years at risk for males vs. 3 for females). In another large study among the Ache people of Paraguay \citep{AcheMortality}, accidents and external warfare account together for 37\% of male deaths vs. 18\% of female deaths. 

However, if a higher male than female death rate is mainly due to violence, it cannot be the ultimate explanation for a TMRCA ratio higher than 1, as higher levels of lethal violence among males pinpoints sex-specific behaviors that require a more profound explanation. This is discussed further in Section \ref{summary}.

As suggested in \citet{Hammer2008}, another mechanism that could be responsible for the two-to-one female-male TMRCA ratio in humans is a sex-biased bottleneck at some time in the past, during which more males than females died. It is possible to get a TMRCA ratio higher than 1 by introducing a sex-biased bottleneck in our ABM as follows: we start a simulation with the usual carrying capacities $K_{F}=K_{M}=75$ and impose for instance $K_{M}=2$ during 300 time steps ($>G$) near the end of the simulation (e.g. at time step $1.7 \times 10^{6}$ within a simulation of total duration $2 \times 10^{6}$ time steps), and then restore $K_{F}=K_{M}=75$ for the remaining time steps (results not shown). An interesting constraint is that the duration of the bottleneck has to be longer than the pregnancy length. Otherwise, many unrelated males are born shortly after the bottleneck from females who were pregnant before the bottleneck started, which considerably lowers its effect on paternal ancestry lines. Therefore a sex-biased bottleneck would necessitate that males and females facing the same hostile environmental conditions have very different survival rates in a consistent way over several months or years. Warfare with indigenous tribes, being attacked by or hunting unknown animals, toxic food ingestions or simply accidents during a long-distance migration through new territories could be explanations for a one-off (or repeated) sex-biased bottleneck, but again, such sex-biased exposures to death causes would require a deepest explanation, as discussed in Section \ref{summary}.

\subsection{Other mechanisms} \label{others}

We tested the effect of sex-specific maturity ages, namely females reaching sexual maturity before males, so as to reflect findings of anthropological studies showing that the age at first reproduction (AFR) tends to be younger for females than for males \citep{Helgason, TremblayVezina}. As explained in \citet{Tang}, increasing the male generation length in a genetic model would decrease the per-year NRY mutation rate, so that genetic distance would be larger for mtDNA than for NRY and coalescent models would infer a larger female than male TMRCA (counted in years). However, in our non-genetic model, increasing the male AFR actually reduces slightly the number of children a male can have in his lifetime and the male variance in reproductive success is reduced as well. The average TMRCA ratio remains close to 1 under this condition in all mating systems with a homogeneous male population (results not shown).

We also tested the effect of letting menopausal females continue to participate in the mating interactions, thus mating without ever getting pregnant anymore. In the polygynandry mode, this uniformly affects all males in making them susceptible to `losing their turn' by picking menopausal females. It thus slightly reduces the male variance in reproductive success, so the average TMRCA ratio remains close to 1. In the polygyny mode, having a menopausal wife decreases a male's adjusted fitness $F_{i}$ without bringing him the benefit of having additional children from her. Again, this affects all males uniformly and does not make the average TMRCA ratio rise above 1 (results not shown).

\subsection{Female and male ancestors}

Baumeister's interpretation of the observed TMRCA ratio is that `today's human population is descended from twice as many women as men' \citep{Baumeister}. The claims associated with this interpretation need clarification. First, Baumeister makes use of two concepts as if they were equivalent, which is not the case: (i) breeding individuals (`throughout the entire history of the human race, maybe 80\% of women but only 40\% of men reproduced') and (ii) ancestors as determined from the presently living individuals (`Today's human population is descended from twice as many women as men', `Most men who ever lived did not have descendants who are alive today. Their lines were dead ends.'). As the TMRCA ratio reflects characteristics of ancestry lines, we retain the concept relating to the number of ancestors.

Second, it is unclear whether Baumeister counts our ancestors along all possible ancestry lines or only along all-female and all-male lines. In the latter case, let us call the corresponding ancestors `mtDNA-ancestors' and `NRY-ancestors'. Consider a man from a past generation who had no sons, but had a daughter whose descendant line reaches someone from the present population (us). Then this man is not one of our NRY-ancestors, but he is an ancestor in the genealogical sense, and he potentially transmitted some of his DNA to us, although not including his NRY. In other words, the reproductive success of such a man is not to be overlooked. Because mtDNA and NRY are just two of a vast number of loci on the human genome, this example illustrates the fact that the numbers of our mtDNA- and NRY-ancestors have little significance, if any, in terms of reproductive success and genetic transmission. Thus, if Baumeister meant that we have twice as many mtDNA- as NRY-ancestors, no evolutionary behavioral conclusion can actually be drawn for that assumption.

On the other hand, the TMRCA ratio informs us about all-female and all-male ancestry lines, and says nothing about our ancestors along all possible ancestry lines. Thus, if Baumeister meant that we have twice as many female as male ancestors in the general sense, this quantitative claim cannot be justified on the basis of the observed TMRCA ratio. It is our purpose to test this guess in the present section.

To that aim, we count the number of agents whose descendant line along any genealogical path reaches at least one agent of the present generation. We call these agents `ancestors' and categorize them according to their gender. Let us stress that the lines of descent can follow any path (e.g. female-male-male-female-etc.), not only matrilines and patrilines, and the gender of an ancestor's descendant(s) belonging to the present generation does not play any role. This means that from the viewpoint of the present generation, non-ancestors are genealogical dead ends, or extinct individuals. The numbers of female and male ancestors give a fair measure of the transmission success of each gender, in a genealogical and potentially genetic sense. The results are shown in Table \ref{TableAncestors}.

The ratio of the number of female ancestors over the number of male ancestors, $r_{A}$ in Table \ref{TableAncestors}, is approximately 1.4 in the modes that maximize the probability that the TMRCA ratio lies between 1.5 and 2.5. The standard deviations of all values shown in Table \ref{TableAncestors} are very small, so only that of $r_{A}$ is shown. This ratio $r_{A} \simeq 1.4$ translates into $\approx$ 20\% of the males becoming ancestors ($\approx$ 42\% of the mature males), compared with $\approx$27\% of the females ($\approx$ 59\% of the mature females). In other words, $\approx$20\% of all males born during a simulation have a descent line that reaches the present generation, as compared with $\approx$27\% of all females. These values also hold within each generation, on average. For instance, approximately $\approx$20\% of the males within each generation (except for the generations that are closest to the present generation) became ancestors of the present generation in the sense given above.

In summary, Baumeister's guess of a female-male ancestors ratio of 2 is of the right order, albeit probably
overestimated by 30\% according to our simulations. Yet, this does not prove that the observed TMRCA ratio $r_{F/M}=T_{F}/T_{M} \simeq 2$ could not be found for quite different female-male ancestors ratios under different simulation conditions. Because, as already stressed, there is no logical link between the female-male ancestors ratio and the TMRCA ratio, the factor of two for the later does not translate into a prediction for the former. We can state that, under the conditions in our ABM that make the TMRCA ratio most likely to be between 1.5 and 2.5, the present generation has 1.4 times as many female as male ancestors. This supports qualitatively the thesis defended by \citet{Baumeister}.

\begin{table*}
{\renewcommand{\arraystretch}{1.5}
\begin{center}
\begin{tabular}{|c|c|c|c|c|c|c|c|}
\hline
Condition & $r_{0}$ & $r_{A}$ & $\sigma(r_{A})$ & $r_{A}^{F}$ & $r_{A}^{M}$ & $r_{A}^{F,mat}$ & $r_{A}^{M,mat}$ \\
\hline
\hline
Polygynandry & & & & & & & \\ 
HMP & 1.0 & 1.13 & 0.01 & 30\% & 26\% & 64\% & 57\%\\
\hline
Polygyny & & & & & & & \\ 
HMP, WREM, DWC & 1.0 & 1.05 & 0.01 & 31\% & 30\% & 67\% & 64\%\\ 
\hline
\hline
Polygynandry & & & & & & & \\ 
$p_{\alpha}=0.12$, $f_{\beta}=0.12$ & 1.0 & 1.38 & 0.02 & 27\% & 20\% & 59\% & 43\% \\ 
\hline
Polygyny & & & & & & & \\ 
WREM, DWC & & & & & & & \\ 
$p_{\alpha}=0.4$, $f_{\beta}=0.2$ & 1.0 & 1.42 & 0.01 & 27\% & 19\% & 58\% & 41\%\\ 
\hline
\end{tabular}
\end{center}}

\caption{Characteristics of the numbers of female and male ancestors of the present generations in four modes. Each row gives the results for one specific simulation. $r_{0}$ is the ratio of the number of females over the number of males who were born during the simulation. It is shown as a control and should always be very close to 1. $r_{A}$ is the ratio of the number of female ancestors over the number of male ancestors; $\sigma(r_{A})$ is the standard deviation of this ratio. $r_{A}^{F}$ is the ratio of the number of female ancestors over the number of females born during the simulation. In other words, it is the fraction of females who became ancestors. $r_{A}^{M}$ is the fraction of males who became ancestors. $r_{A}^{F,mat}$ and $r_{A}^{M,mat}$ are, resp., the fractions of mature females and mature males who became ancestors. HMP stands for Homogeneous Male Population. The following abbreviations are explained in Section \ref{SecPolygyny}: WREM $=$ Widows Remarry, DWC $=$ Deceased Wives Count. $p_{\alpha}$ is the proportion of alpha males; $f_{\beta}$ is the fitness of beta males. Parameters are given in Expression \ref{parameters}. Averages are obtained over 500 simulations with approximately 44'000 agents each.}

\label{TableAncestors}
\end{table*}

\section{Summary of results} \label{summary}

The ABM that we have introduced and studied is formulated in a way that allows us to determine exactly the female and male TMRCAs under many different conditions: two fundamentally different mating systems (polygynandry and polygyny) with several variants each and in the presence (or absence) of sex-biased migration. In our ABM, each male has a fitness $f \in \left[0,1\right]$. Depending on the mating mode, this fitness acts as the probability of mating or getting a wife. While alpha males have a fitness $f_{\alpha}=1$, beta males have a fitness $f_{\beta}$ with $0<f_{\beta}<1$ (all beta males have the same fitness). If all males have the same fitness (all-alphas or all-betas), the male population is homogeneous. A male population consisting of both alpha and beta males is heterogeneous.

We find that the mating system as such has little influence on the TMRCA ratio $r_{F/M}=T_{F}/T_{M}$. The most important finding of our study is that homogenous and heterogeneous male populations lead to two very different regimes, the latter being more likely than the former to be responsible for the observed TMRCA ratio of about 2 found by genetic studies (mtDNA-based TMRCA: \citet{Tang, Ingman, Cann}, NRY-based TMRCA: \citet{Tang, Pritchard, Thomson, Hammer2002}). Indeed, the only condition clearly able to account for a TMRCA ratio of 2 is a heterogeneous male population, whereby some males are alphas and others are betas. This implies an important level of competition among males in order to be recognized as an alpha or appear like one.  In order to reproduce the empirical value $r_{F/M} \simeq 2$, our simulations show that less than half of the males can be alphas. In addition, betas can have at most half the fitness of alphas.

Another finding of the present study concerns the hypothesis made by \citet{Baumeister}, namely that the observed TMRCA ratio $r_{F/M} \simeq 2$ means that we have twice as many female as male ancestors. We define an `ancestor' as an individual whose descendant line along any genealogical path reaches at least one agent from the present generation. We find that in the modes maximizing the probability of the TMRCA ratio to be between 1.5 and 2.5, the present generation has approximately 1.4 times as many female as male ancestors. Therefore, our results support Baumeister's interpretation to a certain degree, and at the same time the important behavioral consequences he develops in \citep{Baumeister} regarding higher male than female competitiveness and risk taking.

Because of the life cycle parameters input into our ABM concerning pregnancy and lactation times, males always have a higher variance in reproductive success than females. In the polygynandry mode with homogeneous male population for instance, the standard deviation of the number of children per mature female is 3, compared with 4.2 for males (for an average number of children per mature agent of either sex of 4.3). In addition, the forward-looking process of genealogical tree building exhibits a large stochasticity (see $\sigma(r_{F/M})$ in Table \ref{TableTMRCA}). These two facts contribute in creating a stretched $r_{F/M}$ distribution with values that can rise above 1 even with a homogeneous male population.

For example, in a polygyny variant (DWC/WREM) with homogeneous male population, we obtain an average $\bar{r}_{F/M}=1.5$, a median of 1.2, 9\% of the simulations with $r_{F/M}<0.5$, 52\% with $0.5<r_{F/M}>1.5$, 24\% with $1.5 < r_{F/M} < 2.5$ and the remaining 15\% with $r_{F/M}>2.5$ (over a total of 1000 simulations). However, the contrast with a heterogeneous male population is striking. In the same polygyny variant with 40\% alphas males and betas' fitness five times smaller than alphas', we obtain $\bar{r}_{F/M}=2.9$, a median of 2.3, 1\% of the simulations with $r_{F/M}<0.5$, 26\% with $0.5<r_{F/M}<1.5$, 29\% with $1.5 < r_{F/M} < 2.5$ and 43\% with $r_{F/M}>2.5$ (over 1000 simulations).

Hence, the probability of finding a TMRCA ratio between 0.5 and 1.5 is reduced about twofold with a heterogeneous male population as compared to a homogeneous population. Because of the high stochasticity of the process and the relatively small population size (150 individuals in our ABM), increasing the average TMRCA ratio goes together with flattening the TMRCA ratio distribution. This leads to at most 30\% of the simulations with $1.5 < r_{F/M} < 2.5$ under the conditions tested in our model.

Sex-biased migration in our two-deme island model is able to enhance the TMRCA ratio significantly, but only under the quite unrealistic condition of strictly zero male dispersal. In reality, modern foragers show a fluid multilocal pattern of residence, whereby couples alternate living with the wife's and the husband's kin \citep{MarloweMigr}. From an ethnological point of view, the male and female migration rates are thus expected to be similar among modern humans before the transition to agriculture \citep{Wilkins}, and not comply with the condition found in our migration model. However, it should be noted that this migration model is very simplified, as agents can move only once in a lifetime and only between two island-like demes without any concept of isolation by distance or population split. Yet a genetic study showed that global NRY and mtDNA patterns are not shaped by female sex-biased dispersal \citep{Hammer2004mig}, although human migration in general plays a complex and important role in shaping patterns of genetic variation \citep{Hammer2008, Hammer2010}.

Another mechanism that can yield an average TMRCA ratio above 1 is a higher male than female death rate either over the whole simulation duration or over a given number of time steps, corresponding to a sex-biased bottleneck. However, unequal death rates sustained over the time necessary to induce unequal TMRCAs would unlikely be caused by chance alone. They would need to rest on an existing gender discrepancy in the susceptibility to the main cause of death. Although modern medical studies show that males are more susceptible than females to diseases \citep{Hon}, we are unaware of any disease occurring in hunter-gatherer societies that would affect males more than females at the level necessary to induce a sex-biased TMRCA ratio (according to our results, such a disease would have to keep alive at least 1.5 times more females than males). On the other hand, any form of violence or accidents seems a more likely sex-specific death cause as it correlates with anthropological findings \citep{HiwiMortality, AcheMortality}. High levels of violence and accidents among males would mean that there is an underlying male-male competitive drive to become skilled and/or to be less exposed to hazards, or to be perceived as either strong or socially integrated, and so on. 

Putting all results together, the underlying explanation that appears to be favored by our results is that unequal biological costs of reproduction is the root cause of unequal TMRCAs, this cost asymmetry cascading into female choosy selection \citep{Baumeister, Wicked}. Female choice causes male-male competition and male's honest signaling, which takes the form of risk-taking behavior \citep{Zahavi, Diamond}. This further cascades into higher male than female death rates through risky signaling and results in a smaller male than female breeding population, both because females select a subset of males for reproduction and because of male's higher death rate. Mild polygyny appears as a by-product of this chain mechanism, favored by female selection and male deaths.

Tables \ref{TableTMRCA} and \ref{TableRS} compile the characteristics of the distribution of the TMRCA ratio and of the distribution of reproductive success under all conditions.

\begin{table*}[p] 
{\renewcommand{\arraystretch}{1.5}
\begin{tabular}{|c|c|c|c|c|c|}
\hline
Condition & $\bar{r}_{F/M}$ & $\sigma(r_{F/M})$ & $\mu_{1/2}(r_{F/M})$ & $0.5 < r_{F/M} < 1.5$ & $1.5 < r_{F/M} < 2.5$ \\
\hline
\hline
Polygynandry & & & & & \\ 
HMP & 1.2 & 0.8 & 1.0 & 58\% & 20\% \\
\hline
Polygyny & & & & & \\ 
HMP, WnoREM, DWnoC & 0.8 & 0.6 & 0.6 & 54\% & 7\% \\ 
\hline
Polygyny & & & & & \\ 
HMP, WnoREM, DWC & 1.1 & 0.9 & 0.9 & 59\% & 16\% \\ 
\hline
Polygyny & & & & & \\ 
HMP, WREM, DWnoC & 1.2 & 0.9 & 0.9 & 55\% & 17\% \\ 
\hline
Polygyny & & & & & \\ 
HMP, WREM, DWC & 1.5 & 1.1 & 1.2 & 52\% & 24\% \\ 
\hline
\hline
Polygynandry & & & & & \\ 
$p_{\alpha}=0.12$, $f_{\beta}=0.12$ & 2.8 & 2 & 2.2 & 26\% & 29\% \\ 
\hline
Polygyny & & & & & \\ 
WREM, DWC & & & & & \\ 
$p_{\alpha}=0.4$, $f_{\beta}=0.2$ & 2.9 & 2.2 & 2.3 & 26\% & 29\% \\ 
\hline
Sex-biased migration & & & & & \\ 
Polygynandry, HMP & & & & & \\ 
$m_{F}=0.1$, $m_{M}=0$ & 2.1 & 1.4 & 1.7 & 37\% & 32\% \\
\hline
Unequal death rates & & & & & \\ 
Polygynandry, HMP & & & & & \\ 
$K_{F}=120$, $K_{M}=75$ & 2.9 & 2.1 & 2.3 & 23\% & 30\% \\
\hline
\end{tabular}}

\caption{Characteristics of the distribution of the TMRCA ratio $r_{F/M}=T_{F}/T_{M}$ for all modes. $\sigma$ and $\mu_{1/2}$ denote respectively the standard deviation and the median. Averages are denoted by bars over the variables. $a < r_{F/M} < b$ gives the proportion of the simulations with a TMRCA ratio comprised between a and b. HMP stands for Homogeneous Male Population. The following abbreviations are explained in Section \ref{SecPolygyny}: WREM $=$ Widows Remarry, WnoREM $=$ Widows don't Remarry, DWC $=$ Deceased Wives Count, DWnoC $=$ Deceased Wives don't Count. $p_{\alpha}$ is the proportion of alpha males; $f_{\beta}$ is the fitness of beta males; $m_{F}$ and $m_{M}$ are the female and male migration rates; $K_{F}$ and $K_{M}$ are the female and male carrying capacities. Parameters are given in Expression \ref{parameters}, except for $K_{F}$, $K_{M}$ and the number of iterations (3 million) in the unequal death rates mode, and the number of iterations (2.5 million) in the migration mode. Averages are obtained over 1000 simulations.
}
\label{TableTMRCA}

\end{table*} 

\begin{table*}[p]
{\renewcommand{\arraystretch}{1.5}
\begin{tabular}{|c|c|c|c|c|c|c|c|c|c|}
\hline
Condition & $\bar{k}_{F}$ & $\bar{k}_{M}$ & $\overline{\sigma}(k_{F})$ & $\overline{\sigma}(k_{M})$ & $\overline{\max(k_{F})}$ & $\overline{\max(k_{M})}$ & $\overline{w}_{F}$ & $\overline{w}_{M}$ \\
\hline
\hline
1. Polygynandry & & & & & & & &  \\ 
HMP & 4.3 & 4.3 & 3 & 4.2 & 9 & 26 & 5\% & 17\%  \\
\hline
Polygyny & & & & & & & &  \\ 
HMP, WnoREM, DWnoC & 3.4 & 3.4 & 2.6 & 3.4 & 9 & 16 & 2\% & 29\%  \\ 
\hline
Polygyny & & & & & & & &  \\ 
HMP, WnoREM, DWC & 3.5 & 3.5 & 2.5 & 3 & 9 & 19 & 2\% & 10\%  \\
\hline
Polygyny & & & & & & & & \\ 
HMP, WREM, DWnoC & 4.3 & 4.3 & 2.9 & 3.9 & 9 & 22 & 3\% & 11\%  \\ 
\hline
Polygyny & & & & & & & &  \\ 
HMP, WREM, DWC & 4.3 & 4.3 & 2.9 & 3.6 & 9 & 24 & 3\% & 7\%  \\ 
\hline
\hline
Polygynandry & & & & & & & &  \\ 
$p_{\alpha}=0.12$, $f_{\beta}=0.12$ & 4.3 & 4.3 & 3 & 8 & 9 & 82 & 4\% & 25\%  \\ 
\hline
Polygyny & & & & & & & &  \\ 
WREM, DWC & & & & & & & &  \\ 
$p_{\alpha}=0.4$, $f_{\beta}=0.2$ & 4.3 & 4.3 & 2.9 & 7.8 & 9 & 47 & 3\% & 58\%  \\ 
\hline
Sex-biased migration & & & & & & & &  \\ 
Polygynandry, HMP & & & & & & & &  \\ 
$m_{F}=0.1$, $m_{M}=0$  & 4.3 & 4.3 & 3 & 4.2 & 9 & 26 & 4\% & 17\%  \\ 
\hline
Unequal death rates & & & & & & & &  \\ 
Polygynandry, HMP & & & & & & & &  \\ 
$K_{F}=120$, $K_{M}=75$ & 4.3 & 7 & 3 & 7.2 & 9 & 51 & 4\% & 12\%  \\
\hline
\end{tabular}}
\caption{Characteristics of the distributions of male and female reproductive success for mature agents in all modes. The number of children per mature female and male are denoted by $k_{F}$ and $k_{M}$, resp. Averages are denoted by bars over the variables. $\sigma$ denotes the standard deviation. $w_{F}$ and $w_{M}$ are the proportions of childless mature females and males. HMP stands for Homogeneous Male Population. The following abbreviations are explained in Section \ref{SecPolygyny}: WREM $=$ Widows Remarry, WnoREM $=$ Widows don't Remarry, DWC $=$ Deceased Wives Count, DWnoC $=$ Deceased Wives don't Count. $p_{\alpha}$ is the proportion of alpha males; $f_{\beta}$ is the fitness of beta males; $m_{F}$ and $m_{M}$ are the female and male migration rates; $K_{F}$ and $K_{M}$ are the female and male carrying capacities. Parameters are given in Expression \ref{parameters}, except for $K_{F}$, $K_{M}$ and the number of iterations (3 million) in the unequal death rates mode, and the number of iterations (2.5 million) in the sex-biased migration mode. Averages are obtained over 1000 simulations with approximately 44'000 agents each, except for the unequal death rates and the sex-biased migration modes that simulate approximately 106'000 and 55'000 agents per simulation, resp.}
\label{TableRS}
\end{table*}

\section{Discussion} \label{discussion}

The empirical data to which we compare our model simulations refers to a unique human evolutionary history, which corresponds to a single realization of a complex stochastic process. The comparison between our simulations and empirical data is thus not straightforward. However, our numerical construction of the distribution of the TMRCA ratio $r_{F/M}=T_{F}/T_{M}$ allows us to estimate the likelihood of each model using the standard statistical prescription of estimating the likelihood of the data given the model. In other words, the fraction of simulations for different ranges of TMRCA ratios reported in the above figures and tables provide the corresponding likelihood of the model with respect to the empirical evidence (i.e. the observed TMRCA ratio $r_{F/M} \simeq 2$).

Overall, our results support the hypothesis that we are descended from males who were highly successful in terms of reproductive output in a highly male-male competitive context, while females were exposed to a much lower level of female-female competition. This is not to say that there is no female-female competition, but it does usually not take the same form and does not lead to such severe life-threatening consequences as male-male competition \citep{Campbell}. Indeed, male choice (resulting in female-female competition) can be expected in humans because of high paternal investment \citep{Geary} and is evidenced by some female physical characteristics \citep{Jasienska}. Actually, the introduction of male choice in our ABM would reduce the female breeding population and thus the female TMRCA. Therefore, taking male choice into account in our ABM would have required even higher levels of male population heterogeneity in order to reproduce the observed two-to-one female-male TMRCA ratio.

We would like to stress that, by highlighting male-male competition, we are not dismissing the essential role of cooperation in the evolutionary history of humans. It seems to us that the two concepts are not exclusive. For instance, cooperative hunting and sharing offer an opportunity for reputation building, which can be positively correlated with mating success \citep{hunting}. Communal male hunting can thus integrate an element of competition for being the best hunter. Furthermore, within-group cooperation might be maintained or enhanced by between-groups competition, although this is debated (\citet{Bowles} argues in this sense but is contradicted by \citet{Langergraber}). Finally, one of us implemented an ABM of a public good game with altruistic punishment and found that cooperation can thrive among selfish disadvantageous inequity averse agents \citep{Hetzer}, a conclusion also supported by game theoretical calculations \citep{Darcetsor,Hetzer2}. This suggests that competitive agents still find it advantageous to cooperate in order to achieve their goals.

That males have inherited a higher propensity to compete than females is reflected by recent psychological and sociological studies. In a laboratory experiment where participants were given the choice between a competitive tournament and a non-competitive reward, men chose to enter the competitive environment twice as much as women, although there were no gender differences in performance \citep{Niederle2007}. Rather than sex-specific risk or feedback aversion, this result was caused by men being more overconfident than women and preferring competition. A milestone study of household investments showed that men are more overconfident than women \citep{BarberOdean}. This resulted in men trading more and consequently hurting their performance more than women. In contrast, a recent chinese study found that women may in fact be more competitive than men (and/or universities may sort genders differently by competitive attitude) \citep{tigerwomen}. This study used an all-pay auction experiment where winning only depends upon willingness to pay, in order to rule out ability and confidence confounds of real tests used in prior studies.

One domain of modern activity where differences in gender attitudes may be prominent is the financial area. It is well documented that men trade more than women and this consequently hurts their performance 
as they underperform on average their female peers. The idea is now emerging that the overwhelming majority of young men on trading floors might play a role in letting financial bubbles and crashes go out of control \citep{Coates2010}. In the light of the results presented in the present paper, we propose that a high male-male competition for reproductive success that has been permeating the history of modern humans (200'000 years ago to recent times) has contributed through gene-culture coevolution to create gender competitiveness-related differences. These differences are not fully understood yet and we believe that investigating their evolutionary causes can help deal with the possible negative consequences associated with a maladaptation to the modern era. These considerations raise the question of how to adjust our cultures and/or design society rules in order to match better the evolutionary-based inherited human traits described above and the requirements of the modern ages.

Our ABM could be further improved in future work by adding genetics. In particular, agents could be endowed with genetic sequences that are inherited, undergo mutations, determine agent fitness and are thus under selection pressure. A more realistic population structure could also be implemented in order to investigate further the effects of migration. Finally, after integrating genetics, it would be highly interesting to see whether existing softwares that infer TMRCAs using DNA samples are able to find the TMRCAs realized by our ABM with a reasonable accuracy under different mating systems and life cycle characteristics, when given only a sample of genetic sequences.

\section*{Acknowledgments}

For fruitful discussions and valuable feedbacks, we would like to thank Philipp Becker and Pirmin Nietlisbach, Fr\'ed\'eric Guillaume, Natasha Arora and Alexander Nater, Tanja Stadler, Moritz Hetzer and Christine Sadeghi. This work was partly supported by the Swiss National Science Foundation.

\bibliographystyle{model2-names}
\bibliography{ancestors_biblio} 

\end{document}